\documentclass[preprint,prd,aps,showpacs,showkeys,onecolumn,english]{revtex4}
\usepackage{amsmath,amssymb} % Enhanced mathematics
\usepackage[dvips]{graphicx} % Include figure files
\usepackage{graphics}
\usepackage{graphicx}
\usepackage{epsfig}
\usepackage{slashed}
\usepackage[english]{babel}
\usepackage{amsmath}
\usepackage{amssymb}
\usepackage{graphics}
\usepackage{graphicx}
\usepackage{epsfig}
\usepackage{slashed}
\usepackage{subfigure}	% Multiple figures on one plot
\usepackage[export]{adjustbox}
\newcommand\nn{\nonumber}
\newcommand\ba{\begin{eqnarray}}
\newcommand\ea{\end{eqnarray}}
\newcommand{\br}[1]{\left( #1 \right)}

\newcommand{\GeV}{~\mbox{GeV}}
\newcommand{\MeV}{~\mbox{MeV}}

\newcommand{\eV} {~\mbox{eV}}

\newcommand{\Tr}{\mbox{Tr}}

\begin{document}
\title{Cross section of the process $e^+ + e^- \to \Xi^0 + \bar{\Xi}^0$ in the vicinity of charmonium $\psi(3770)$
and in the charmonium (-like) state including the $D$-meson loop and three-gluon contributions}
\author{Azad I.~Ahmadov$^{a,b}$ \footnote{E-mail: ahmadov@theor.jinr.ru}}
\affiliation{$^{a}$ Institute of Physics,
Ministry of Science and Education, H.Javid avenue, 131, AZ1143 Baku, Azerbaijan}
\affiliation{$^{b}$ Bogoliubov Laboratory of Theoretical Physics,
JINR, Dubna, 141980 Russia}

\date{\today}

\begin{abstract}
In the present work, I explore the production of $\Xi^0 \bar{\Xi}^0$ using $e^+e^-$ collision
data at ten center-of-mass energies between 3.51 and 4.95 GeV collected with the BESIII detector
at the BEPCII collider and corresponding to an integrated luminosity of 30 $\rm{fb}^{-1}$ to study the
structure of baryons.
The data collected by the BESIII detector are useful for the study of XYZ states, and this collaboration
continues the exploration of these exotic charmoniumlike states.
At higher center-of-mass energies the baryon-antibaryon pair production in electron-positron annihilation provides a powerful tool.
In this paper, I investigate a hyperon pair produced in the electron-positron annihilation reactions,
and I present the phenomenological results for the process using the BESIII detector at the BEPCII Collider.
In the reaction $e^+e^- \to \Xi^0 \bar{\Xi}^0$, I consider two different contributions:
one from the $D$-meson loop and the other from the three-gluon charmonium annihilation.
In this process I calculate the total cross section including the contributions of the $D$-meson loop and
three-gluon loops as well as the interference of all diagrams.
In the case of a purely electromagnetic mechanism for these contributions, large relative phases are generated.
For a large momentum transferred region I get as a byproduct a fit of the electromagnetic form factor of the $\Xi^0$ hyperon.
In the present work, in addition to the $\psi(3770)$ charmonium, I also take into account the contributions of
other charmonium (-like) states, such as $\psi(4040)$, \,\,$\psi(4160)$, \,\,$Y(4230)$,\,\,$Y(4360)$,\,\,$\psi(4415)$, and $Y(4660)$.
On the whole my results are in good agreement with the available experimental data.

\vspace*{0.5cm}

\noindent
\end{abstract}

\maketitle

% ======================================================================
\section{Introduction}
\label{Introduction}
% ======================================================================

The study of the processes of annihilation of colliding electron-positron beams and inelastic
lepton-nucleon scattering has played a significant role in the development of modern elementary particle physics.
Most of the phenomena and whole families of hadron resonances $\psi$ and $\Upsilon$ were discovered precisely
in these experiments, which play an important role in understanding the nature of fundamental interactions and
the structure of elementary particles.
Particle physics, using quarks and leptons and their interactions, tries to explain the diversity of the Universe.
One of the main sources of fundamental information in elementary particle physics is experiments on electron-positron interactions.

The production of hadrons with a pair of baryons or strange mesons in the final state in the
$e^+e^-$ annihilation process at high energies has been one of the main tasks of experimental study on the basic
structure of the Standard Model for many years for several reasons.
First, they give a significant contribution to the total cross section of $e^+e^-$ hadrons.
Second, timelike electromagnetic form factors (EMFFs) can be derived from these reactions.
Also, when studying these reactions, new states can be discovered.
Experiments with $e^+e^-$ interactions have given several interesting results in the vector meson region
on the annihilation cross section as well as some preliminary results in the region of several $\text{GeV}$.

Note that after the discovery of hyperons and the study of the mechanisms of their formation, decay processes
became of current interest in hadron physics.

It is important to note that below the threshold of $D\bar{D}$, the spectrum has been well studied experimentally and theoretically.
Therefore, vector states can be easily studied with high accuracy in $e^+e^-$ annihilation in resonance production.
Namely, the BESIII experiment plays a significant role and provides a good opportunity to study these states due to the
high statistical data samples.

Using $e^+e^-$ annihilation at electron-positron linear colliders data taken by the BESIII detector,
several searches for charmoniumlike states were performed, that is, $\psi(4040),\,\psi(4160),\,Y (4230),\,Y (4360),
\,\psi(4415),\,Y (4660)$ and others.

A number of new hadron states, including charmoniumlike XYZ states, have generated extensive
discussions for the study of the exotic hadron family \cite{Liu2014,Chen2016,Chen2017,Guo2018,Liu2019,Brambilla2020,Chen2023}.
Among these new states were observed charmoniumlike XYZ states, that have quantum numbers $J^{PC} = 1^{--}$, $Y(4260)$ \cite{BABAR1,BELLE1,He},
$Y(4360)$, and $Y(4660)$ \cite{BABAR2,BELLE2,BABAR3,BELLE3} and are formed as a result of the $e^+e^-$ annihilation process, which are called $Y$ states.
Thus, charmonium production with energies above the threshold is of great theoretical interest due to its richness in $c\bar {c}$ states.
It is necessary to note that experimental measurements of charmonium decays may become an ideal laboratory for studying
the dynamics of strong interaction physics and testing various aspects of quantum chromodynamics (QCD) \cite{Ablikim2019,Brambilla2004}.
Thus, after the experimental discovery of entirely new exotic hadronic states in this region, significant progress
can be expected in charmonium spectroscopy above the open-charm threshold.
At this time, BESIII is the only experiment that collects data using $e^+e^-$ collisions in the field of charmonium.

It is understood that hyperons are SU(3)-flavor-octet partners of nucleons that contain strange quarks and offer
some extra dimensions for exploring nucleon structures \cite{Eichmann,Ramalho1,Gross,Ramalho2,Ramalho3}.

In this paper, I use timelike EMFFs because they provide fundamental information and describe the modifications of the
pointlike photon-hadron vertex due to the internal structure and dynamics of hadrons,
because, EMFFs are used to understand the effects of QCD in hadron resonances, which parameterize the internal structure
and dynamics of hadrons \cite{Geng,Brodsky2,Green,Ablikim2020}.
It should be noted that timelike EMFFs were measured in the processes $e^+e^-\to B\bar{B}$
\cite{Delcourt1979,Antonelli1998,Armstrong1993,Pedlar2005,Ablikim2018} (where $B$ denotes the ground baryon state with spin 1/2).

The electromagnetic structure of hadrons was studied in the original papers of Hofstaedter and McAllister \cite{Hofstadter1,Hofstadter2},
which remains an open and interesting object of research in high-energy physics.

In papers \cite{Buchmann1,Buchmann2}, an approach to obtaining additional information on the geometric shape of the nucleon and
the intrinsic quadrupole form factor of the proton was investigated, and it was shown that the form factor of
the neutron charge was an observable manifestation of the intrinsic quadrupole form factor of the nucleon.
In the papers \cite{Buchmann3,Buchmann4}, the results of calculations of the higher electromagnetic multipole moments of baryons
and quadrupole moments of decuplet baryons are presented within the framework of the noncovariant quark model.

It is worth noting that in the study of the initial state radiation process $e^+e^- \to \eta J/{\psi}$, in the Belle experiment,
only the well-known charmonium states $\psi(4040)$ and $\psi(4160)$ \cite{BELLE4} were detected.

It is know that the resonance $\psi(3770)$ decays almost entirely into pure $D\bar{D}$ \cite{DELCO}, and the resonance
$\psi(3770)$ is the lowest-mass resonance of charmonium above the open charm pair formation threshold $D\bar{D}$.

Usually in experimental measurements all charmonium states were observed below the mass threshold for charm pairs
($2m_D$), and their observed spectrum was known to be consistent with the charm-anticharm potential model prediction \cite{Godfrey}.

To analyze the cross section in the $\bar {p} p\to e^+e^-$ and $e^+e^- \to \bar{p}p$ \cite{Haidenbauer} processes
in the near-threshold region, for the proton the effective form factors ($G_E$ and $G_M$ form factors) were used.

Therefore, the cross section of the process of $e^+e^-$ annihilation into hadrons (hyperon pairs)
is described in terms of the EMFFs.

Form factors are analytic functions, and in the timelike region of the $e^+e^-$ annihilation process for every
value of the square of the transferred momentum $q^2$ they are positive and correspond to the square of the
center-of-mass (c.m.) energy, $q^2 = s > 0$.

For the production of a nucleon-antinucleon pair ($\bar{N}N$) in the $e^+e^-$ annihilation process
in the Born approximation with a distorted wave, the effective EMFFs of a proton and a neutron in the timelike region
were studied in \cite{Qin}.

In \cite{Cabibbo1,Cabibbo2} it was suggested that by measuring the cross sections of hadron pair production
in $e^+e^-$ collisions the EMFFs of hadrons can be studied for timelike regions, $q^2>0$.
It is worth noting that in several experimental papers \cite{Ablikim2020,Ablikim1042021,Ablikim972018,Ablikim7352014,
Aubert762007,Lees872013,Lees882013,Ablikim912015}, at large $q^2$ \cite{Lepage1979,Belitsky2003} there is an EMFF in the
timelike region, which is consistent with the simple quark counting rules and the prediction of perturbative
quantum chromodynamics (pQCD).

In this regard, BESIII Experiment collected the largest data sample in the interval corresponding to the integrated
luminosity from 5 to 30 $fb^{-1}$ in the energy range from 2.3864 to 4.95 GeV  to study the XYZ states,
i.e., around the resonances $\psi(3770)$,\,\, $\psi(4040)$, \,\,$\psi(4160)$, \,\,$Y(4230)$,\,\,$Y(4360)$,\,\,$\psi(4415)$, and $Y(4660)$
\cite{Ablikim392015,Ablikim1032021,Ablikim8312022,Ablikim112023,Ablikim052024,Ablikim112024}.

It is necessary to note that the decay process into light hadrons in these charmonium states occurs either via the one-photon
process ($e^+e^-\to \psi \to \gamma^* \to hadrons$) or the three-gluon process ($e^+e^-\to\psi \to ggg \to hadrons$).

According to the Okubo-Zweig-Iizuka (OZI) rule, since the charmonium $\psi(3770)$ lies above the $D\bar{D}$ threshold,
it will therefore predominantly decay to $D\bar{D}$ final states \cite{DELCO,Rapidis1978,Pallin1987}.

In the theory of hadron physics, the resonance of $\psi(3770)$ \cite{PDG} at 3.770 GeV, due to the
richness of $c\bar{c}$ states, turned out to be the only observed structure that exceeds the threshold for the
formation of an open charm pair of $D\bar{D}$ and, therefore, the study of such a resonance is of great interest,
which is one of those important structures in the hadron cross section.

The BESIII detector operating at the BEPCII collider in the energy range from 2.3864 to 4.95 GeV for
different integrated luminosities has so far conducted high-precision studies of a possible threshold increase in the processes
$e^+e^- \to \Sigma^{\pm} \bar{\Sigma}^{\mp}$ \cite{Ablikim052024,Ablikim8142021},
$\Xi^-\bar{\Xi}^+$ \cite{Ablikim1032021,Ablikim112023}, $\Sigma^0 \bar{\Sigma}^0$ \cite{Ablikim8312022} and $\Xi^0 \bar{\Xi}^0$
\cite{Ablikim112024} and also one can understand that measuring the near-threshold pair generation of hyperon production
in the $e^+e^- \to \Xi^0 \bar{\Xi}^0$ process \cite{Ablikim112024} the threshold effect obtained in this way will be useful.

I also would like to note that the process $e^+e^- \to \Xi^0 \bar {\Xi}^0$ was studied both experimentally and also theoretically \cite{Ablikim8202021,Ablikim7702017,Ablikim062022,Dobbs1,Dobbs2,WangCHARM2020}, and indeed, several experiments are of great
importance for the development of high-energy particle physics.
I want to emphasize that the aim of the present paper is to explore the characteristics of the production of
$\Xi^0 \bar {\Xi}^0$ in the reaction $e^+e^- \to \Xi^0 \bar {\Xi}^0$ including, in addition to the one-photon exchange,
also the exchange of several states of charmonium (-like) $\psi(3770),\, \psi(4040),\, \psi(4160),\,Y(4230),\, Y(4360),\,
\psi(4415)$, and $\psi(4660)$.

Experiments at current and future colliders are expected to provide an enormous amount of data and greatly increase
sensitivity to new physics if equally accurate predictions of the Standard Model are available.

\section{Process $e^+e^- \to \Xi^0 \bar{\Xi}^0$ in Born approximation \label{Born}}

In this section, I consider the $\Xi^0 \bar{\Xi}^0$ production at the electron-positron annihilation reactions.
The kinematics for the annihilation reaction is best described in the c.m. system.
This process is presented in the following form:
\ba
e^+(p_+) + e^-(p_-) \to \Xi^0(q_1) + \bar{\Xi}^0(q_2),
\label{process}
\ea
where $p_+$ and $p_-$ are the momenta of the initial state positron and electron, and $q_1$ and $q_2$ are the
momenta of the final state $\Xi^0$ and $\bar{\Xi}^0$ hyperons, respectively.

According to the process \eqref{process}, one can define the Mandelstam invariants as follows:
\ba
s &=&(p_+ + p_-)^2 = (q_1 + q_2)^2;  \notag \\
t &=&(p_+ - q_1)^2 = (p_- - q_2)^2;  \notag \\
u &=&(p_+ - q_2)^2 = (p_- - q_1)^2.
\label{Mand1}
\ea
Using the law of conservation of momentum, the sum of the Mandelstam variables satisfies the following expression:
\ba
s + t + u = 2m_{e}^{2} + 2 M_{\Xi}^2.
\label{3}
\ea
Here $m_{e}$ is the mass of the electron and $M_{\Xi}$ is the $\Xi^0$ hyperon mass.
When performing calculations, the mass of the electron, $m_e$, is neglected.

\begin{figure}
    \centering
    \includegraphics[width=0.40\textwidth]{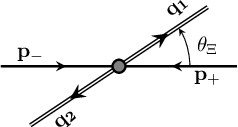}
    \caption{The definition of the scattering angle $\theta_\Xi$ from (\ref{Phi2}) in
    the center-of-mass reference system.}
    \label{fig.MomentaPosition}
\end{figure}

In the center-of-mass frame, the four-momenta of the electron (and positron) in the initial state and hyperons
($\Xi^0$) in the final state can be written in the form
\ba
p_- = \frac{\sqrt{s}}{2}(1,0,0,1),\,\,\,\,p_+ = \frac{\sqrt{s}}{2}(1,0,0,-1),  \nn \\
q_1 = \frac{\sqrt{s}}{2}(1,\beta \sin\theta_\Xi,0, \beta \cos\theta_\Xi),\,\,\,\,\, \nn
q_2 = \frac{\sqrt{s}}{2}(1,-\beta \sin\theta_\Xi,0,-\beta \cos\theta_\Xi). \nn
\ea
In the Born approximation, the differential cross section for the annihilation reaction \eqref{process}
in the center-of-mass system, can be written as:
\ba
d\sigma = \frac{1}{8 s} \sum_{\text{spins}} |{M}|^2 \, d\Phi_2.
\label{CrossSectionGeneralForm}
\ea
Here the sum is performed over all spin states in the matrix element squared.
In the center-of-mass frame element of phase space of the final particles, $d\Phi_2$ has the form,
\ba
&& d\Phi_2 = (2\pi)^4 \delta (p_+ + p_- - q_1 - q_2) \frac{d\bf{q_1}}{(2\pi)^3 2E_1}\frac{d\bf{q_2}}{(2\pi)^3 2E_2} =
\frac{\beta}{16 \pi} \, d\cos\theta_\Xi,
\label{Phi2}
\ea
where $d\Omega = d\phi_{\Xi} d(cos \theta_{\Xi})$ is the solid angle of particle $\Xi$.
$\phi_\Xi$ is the azimuthal angle and $\theta_\Xi$ is the polar angle of the final $\Sigma$-hyperon in the $e^+e^-$
center-of-mass reference frame, i.e., $\theta_\Xi$ is the angle between three-momenta of the directions of the initial
electron $\bf{p_-}$ and the final $\Xi$-hyperon $\bf {q}_1$ (Fig.~\ref{fig.MomentaPosition}).
Here the modulus of three-momenta of the final $\Xi^0$-hyperon (or $\bar {\Xi}^0$-hyperon) can be determined
by the $\delta$-function in the phase space, i.e.,
\ba
&& |{\bf{q}}| \equiv |{\bf{q_1}}| = |{\bf{q_2}}| = \frac{\sqrt{s}}{2}\sqrt {1 - \frac{4M^2_{\Xi}}{s}} = \frac{\sqrt{s}}{2} \beta, \\ \nn
\ea
where $\beta = \sqrt {1 - \frac{4M^2_{\Xi}}{s}}$ is the velocity of the final hyperon in the $e^+e^-$ c.m. system,
$s$ is the $e^+e^-$ total c.m. energy squared, and $M_{\Xi}$ is the mass of the $\Xi^0$-hyperon.

\begin{figure}
	\centering
    \subfigure[]{\includegraphics[width=0.40\textwidth]{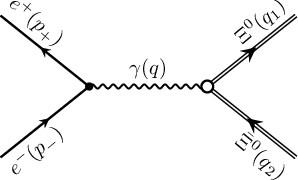}\label{BornDiagram}}
	\hspace{0.05\textwidth}
    \subfigure[]{\includegraphics[width=0.40\textwidth]{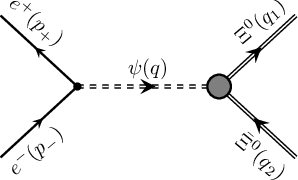}\label{PsiDiagram}}
    \caption{The Feynman diagrams for the $\Xi^0 \bar{\Xi}^0$ hyperon pair production in the $e^+e^-$
     annihilation process corresponding to the one-photon exchange (a) and the intermediate state $\psi(3770)$ charmonium (b).}
    \label{fig.TwoMechanisms}
\end{figure}

Figure~\ref{BornDiagram} shows the Feynman diagrams in the Born approximation for the formation of a hyperon pair $\Xi^0 \bar {\Xi}^0$,
where $e^+e^-$ annihilates \eqref{process} into a virtual photon.

It is known that processes such as \eqref{process}, which are shown in Fig.~\ref{BornDiagram}, are described by a purely electrodynamic diagram.
In the Born approximation, the amplitude for the process ${\mathcal M}_B (e^+e^- \to \Xi^0 \bar{\Xi}^0)$ \eqref{process} has the form:
\ba
{\mathcal M}_B = \frac{1}{s}J^{e\bar{e} \to \gamma}_{\mu}(q)J^{\mu}_{\gamma \to \Xi^0\bar{\Xi}^0}(q)
\label{BornAmplitude}
\ea
where $s = q^2 = (p_+ + p_-)^2 = (q_1 + q_2)^2$ represents the invariant mass squared of the $e^+e^-$ system.
The functions $J^{e\bar{e}(q) \to \gamma}_\mu$ and $J^{\mu}_{\gamma \to \Xi^0\bar{\Xi}^0}(q)$ from \eqref{BornAmplitude} represent
the lepton and hyperon electromagnetic currents, they can be written in the following form:
\ba
   J^{e\bar{e}\to \gamma}_\mu(q) = -e [\bar{v}(p_+) \gamma_\mu u(p_-)],
    \label{BornAmplitude2}
\ea
\ba
   J^{\mu}_{\gamma \to \Xi^0\bar{\Xi}^0}(q) = e [\bar{u}(q_1) \Gamma_\mu(q) v(q_2)],
    \label{BornAmplitude3}
\ea
where $e$ is the modulus electron charge  $e=\sqrt {4\pi\alpha}$ and $\alpha \approx 1/137$ is the quantum electrodynamics (QED)
fine structure constant \cite{PDG}.

To describe the structure of hyperons (baryons) and the exploration of this process in detail, I introduce form factors.
Using the Pauli form factors $F_1$ and $F_2$ for this annihilation process, which is shown in Fig.~\ref{BornDiagram},
one can write the current as $\gamma\Xi^0\bar{\Xi}^0$ $(\gamma B \bar{B})$.
In general, the vertex function $\Gamma_\mu(q)$ describing a photon with $\Xi$ hyperons [Fig.~\ref{BornDiagram}]
is represented in terms of the electromagnetic hyperon form factor:
\ba
	\Gamma_\mu(q) &= F_1(q^2) \gamma_\mu - \frac{F_2(q^2)}{4M_\Xi} (\gamma_\mu \hat{q} - \hat{q} \gamma_\mu),
\label{FF}
\ea
where I use the notation $\hat{q} = q_\nu\gamma^\nu$, and $M_{\Xi}$ and $q$ are the hyperon mass and transferred momentum
in the center-of-mass of the system, respectively.
In \eqref{FF}, the form factors of the $\Xi$-hyperons $F_1(q^2)$ and $F_2(q^2)$ are usually normalized as follows:
$F_1(0)=0$ and $F_2(0)=\mu_{\Xi}$, where $\mu_\Xi$ is the anomalous magnetic moment of the $\Xi$-hyperon.

However, a recent analysis \cite{TomasiG} has shown that the point-like behavior of the hadron near the threshold is not
so unique and the nontrivial structure of the baryon begins to manifest itself already at relatively small $q^2$.
In \cite{Ferroli2012}, it was suggested that in the energy range I am considering, i.e., $\sqrt{s} \sim 3-5~ GeV$,
the approximation of pointlike hadrons is not valid.
Therefore, in the present work, I use the effective form factor G(s).
Besides, without taking into account the form factor, it will be impossible to compare the Born cross section with the experimental data.

Moreover, it is known that in the timelike region there is little statistics and, therefore, the experimental data contain large errors.
At the moment the experiment does not allow us to highlight the electric $G_E$ and magnetic $G_M$ form factors.
Considering this, I use the assumption that $|G_E| = |G_M|$, i.e., $F_2(q^2)=0$.
I want to choose the form factor $F_1$ such that it has a pQCD-inspired form \cite{Lepage1979,Lepage1980} and also takes into account
the running of the QCD coupling constant $\alpha_s$.

After squaring the matrix element in \eqref{BornAmplitude} and further calculating the trace techniques, I then that into
account the form factor $F_1(q^2) = G(q^2)$.
Then for the square of the amplitude I obtain the following expression:
\ba
    \sum_{spins} |{\mathcal M}_B|^2
    &=
    64 \pi^2 \alpha^2 |{G(s)}|^2 ( 2 - \beta^2 \sin^2\theta_\Xi).
    \label{AmplitudeSquare}
\ea
Let us put the formulae for the square of the matrix element \eqref{AmplitudeSquare} and for the phase volume \eqref{Phi2} in
formula \eqref{CrossSectionGeneralForm}. Then, in general, for the differential cross section I get an expression in this form:
\ba
    \frac{d\sigma_B(s)}{d\cos\theta_\Sigma}=\frac{\pi\alpha^2 \beta}{2s}|{G(s)}|^2 (2-\beta^2 \sin^2\theta_\Xi).
    \label{DTotalCrossSectionBorn}
\ea
To obtain an expression for the total cross section, I must perform integrations in \eqref{DTotalCrossSectionBorn}
over all possible scattering angles $d\cos\theta_\Xi = \sin\theta_\Xi\,d\theta_\Xi$.

The limits of integration over the angles $\theta_\Xi$ are defined in this interval $[0 \leq \theta_\Xi \leq \pi]$.

Thus, after performing the integration over the angles $\theta_\Xi$ in formula \eqref{DTotalCrossSectionBorn},
I get an expression for the total cross section in the Born approximation in this form:
\ba
    \sigma_B(s)=\frac{2\pi\alpha^2}{3s} \beta (3-\beta^2) |G(s)|^2.
    \label{TotalCrossSectionBorn}
\ea
Here I use the effective form factor $G(s)$ applied in pQCD \cite{Lepage1979,Lepage1980}, which takes into account the current QCD coupling constant $\alpha_s$:
\ba
	G(s) = \frac{C}{s^2 \log^2\br{s/\Lambda_{\text{QCD}}^2}},
	\label{Formfactor}
\ea
where the constant $C$ is a free parameter that is fitted based on experimental data and the quantity $\Lambda_{\text{QCD}}$ is the QCD scale parameter.
Note that in this process the constant $C$ is fitted for the production of a hyperon pair $\Xi^0 \bar {\Xi}^0$ according to specific experimental data
of the BESIII in the energy range of the corresponding experiment.

\begin{figure}
	\centering
	\includegraphics[width=0.55\textwidth]{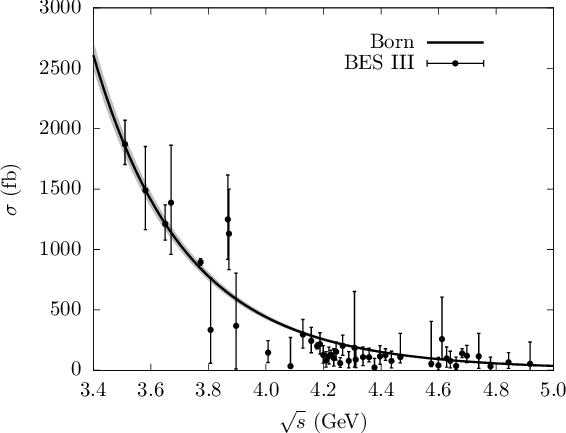}
    \caption{Total cross section distributions of the process $e^+ e^-\to \Xi^0 \bar{\Xi}^0$ as a function of the center-of-mass energy $\sqrt {s}$.
    The Experimental data are from BESIII Collaboration \cite{Ablikim112024}.
    The theoretical curve for the total cross section in the Born approximation (\ref{TotalCrossSectionBorn}) is indicated by black lines.
    Due to errors in fitting of the form factor constant (\ref{eq.C}), curve errors occur.}
    \label{WideRange}
\end{figure}

In this case, for the pair production of $\Xi^0\bar{\Xi}^0$, the constant $C$ is fixed using the whole BESIII measurement range \cite{Ablikim112024},
and the obtained results of the dependence of the total cross section on the center-of-mass energy $\sqrt {s}$ are shown in Fig.~\ref{WideRange}.
In the Born cross section from (\ref{TotalCrossSectionBorn}), if one can accept the quantity $\Lambda_{QCD}$ = 300\,MeV, then for the parameter $C$,
after fitting with respect to these data, one obtains us the following parameter:
\ba
	C = (64.68 \pm 1.3)~\GeV^4.
    \label{eq.C}
\ea
These values of $C$ and $\Lambda_{QCD}$ are further used in our numerical calculations for the electromagnetic form factor
$\Xi$-hyperon (\ref{Formfactor}).
It should be noted that this expression for the form factor $G(s)$ (\ref{Formfactor}) with the parameter $C$ from (\ref{eq.C})
works in the region of relatively large transferred momentum $q^2$.

The threshold near the Coulomb-like amplification factor with many subtle features plays an important role
\cite{Haidenbauer,Amoroso} or it manifests the wave nature of the hyperon stabilization after its exit from the vacuum \cite{TomasiG}.
Therefore, it can not be applied to work near the threshold.

%

% =======================================================================
\section{The quarkonium $\psi(3770)$ intermediate state}
\label{sec.PsiIntermediateState}
% =======================================================================

In this work, for the $\Xi^0 \bar{\Xi}^0$ production in the $e^+e^-$ annihilation process, the main task is to describe the
excitation effect of the charmonium resonance $\psi(3770)$.
It can be seen from Fig.~\ref{WideRange} that the total cross section in the Born approximation of the process \eqref{process},
which includes only the electromagnetic mechanism \eqref{TotalCrossSectionBorn}, in the vicinity of the charmonium
resonance $\psi(3770)$ cannot describe the subtle behavior.
For a detailed study of the process \eqref{process} I need to take into account in this region an additional contribution
to the amplitude, which arises from the diagram with $\psi(3770)$ in the intermediate state (Fig.~\ref{PsiDiagram})
and is enhanced by the Breit-Wigner propagator.
Next, I develop a model (based on our previous calculations \cite{Ahmadov1,Bystritskiy,BA,Ahmadov2,Ahmadov3})
that fully explores the basic mechanism for this process \eqref{process}.
Therefore, in the region of excitation of the charmonium resonance $\psi(3770)$ [$I^G (J^{PC})=0^-(1^{--})$]
in the intermediate state one must compute the contribution of an additional mechanism.
The total amplitude of the process ${\mathcal M}_B (e^+e^- \to \Xi^0 \bar{\Xi}^0)$ \eqref{process} then becomes the sum of two matrix elements:
\ba
\mathcal {M} = \mathcal {M}_B + \mathcal {M}_{\psi},
\label{M}
\ea
where $\mathcal {M}_B$ is the amplitude in the Born approximation (Fig.~\ref{BornDiagram}) from \eqref{BornAmplitude}
for the process \eqref{process} and $\mathcal {M}_{\psi}$ is the amplitude that accounts for the mechanism with
charmonium $\psi(3770)$ for the intermediate state with the enhanced Breit-Wigner factor [Fig.~\ref{PsiDiagram}],
\ba
    M_\psi = \frac{1}{s-M_\psi^2+i M_\psi\,\Gamma_\psi} J^{e\bar{e}\to\psi}_\mu(q) \biggl(g^{\mu\nu} - \frac{q^\mu q^\nu}{M_\psi^2}\biggr)
    J^{\psi \to \Xi^0 \bar{\Xi}^0}_\nu (q),
    \label{Mpsi}
\ea
where $M_{\psi}$ is the mass of the $\psi (3770)$, $M_{\psi}$ = 3773.7 \,MeV,  and $\Gamma_{\psi}$ = 27.2 \,MeV \cite{PDG}
is the total decay width of the $\psi (3770)$ resonance.
The current $J^{e\bar{e}\to\psi}_\mu(q)$ describes the transition of the electron-positron pair to the resonance $\psi(3770)$
and the current $J^{\psi\to \Xi^0 \bar{\Xi}^0}_\nu (q)$ describes the transition of the resonance $\psi(3770)$ to the
hyperon-antihyperon pair, respectively.
The currents in \eqref{Mpsi} must be conserved, that is,
$q^{\mu}J^{e\bar{e}\to\psi}_\mu(q) = q^{\mu}J^{\psi\to \Xi^0\bar{\Xi}^0}_\mu (q) = 0.$
Based on the work \cite{Ahmadov1}, I assume that the vector current $J^{e\bar{e}\to\psi}_\mu(q)$ has also the same
structure as in the case of the photon, that is, $J^{e\bar{e}\to\gamma}_\mu$, and then the vector current, according to
\eqref{BornAmplitude2}, can be written as follows:
\ba
    J^{e\bar{e}\to\psi}_\mu(q) = g_e \, [\bar{v}(p_+) \gamma_\mu u(p_-)],
    \label{Jeepsi}
\ea
where the constant $g_e = F_1^{\Xi^0\bar{\Xi}^0}(M_\psi^2)$ is equal to the value of the form factor
$F_1^{\psi\to \Xi^0\bar{\Xi}^0}(M_\psi^2)$ for the vertex $\psi \to \Xi^0\bar{\Xi}^0$ on the
$\psi(3770)$ mass shell [here I assume that in the case of the Born approximation $F_2^{\psi\to \Xi^0\bar{\Xi}^0}(M_\psi^2) = 0]$.
Note that from \cite{PDG} the total decay width $\psi \to e^+ e^-$ which is $\Gamma_{\psi \to e^+ e^-} = 261~\eV$.
Knowing this, we can compute the constant $g_e$,
\ba
    g_e= \sqrt{\frac{12\pi\Gamma_{\psi\to e^+e^-}}{M_\psi}} = 1.6 \cdot 10^{-3}.
\ea
Based on the work of \cite{Kuraev}, I can neglect the possible imaginary part of the vertex $e\bar{e}\to\psi$,
because in this work it was shown that the imaginary part of the vertex is less than 10~\% of the real part.
Considering this, I do not include this source of error when calculating the statistical uncertainty for the final results.

Using the total matrix element of \eqref{M} and taking into account the contribution of charm to the intermediate state and,
accordingly, the general formula for the total cross section \eqref{CrossSectionGeneralForm} for the total cross section,
I get the following expression:
\ba
	&\sigma \sim |\mathcal {M}|^2 = ||\mathcal {M}_B| + e^{i\phi_{\psi}} |\mathcal {M}_{\psi}||^2
	=\nn\\
	&\qquad = |\mathcal {M}_B|^2 + 2\cos\phi_{\psi} |\mathcal {M}_B| \cdot |\mathcal {M}_{\psi}| + |\mathcal {M}_{\psi}|^2
	\sim \sigma_B + \sigma_{int} + \sigma_\psi,
	\label{TotalCrossSection}
\ea
where $\phi_{\psi}$ is the relative phase between the Born amplitude $\mathcal {M_B}$ and the additional contribution $\mathcal M_\psi$
of the intermediate state $\psi(3770)$ of the charmonium.

Thus, I have to compute the contribution of the charmonium $\sigma_\psi$ to the cross section and the interference contribution
of the Born amplitude in $\sigma_{int}$ with the charmonium amplitude to $\sigma_\psi$.
Knowing the Born section $\sigma_B$ from \eqref{TotalCrossSectionBorn} and the interference contribution of $\sigma_{int}$
with a phase $\phi_{\psi}$, using \eqref{TotalCrossSection} I can calculate the full total cross section
including both contributions, and after that one finds evaluate $\sigma_\psi$ in the following form:
\ba
	\sigma_\psi = \biggl(\frac{\sigma_{int}}{2 \cos\phi_{\psi} \, \sqrt{\sigma_B}} \biggr)^2.
\label{sigmapsi}
\ea
In the present case, based on the general formula \eqref{CrossSectionGeneralForm}, one can calculate the interference
of the Born contribution $\mathcal{M_B}$ with the contribution of the intermediate state $\psi(3770)$ charmonium $\mathcal{M}_\psi$,
which is represented in the following standard form:
\ba
    d\sigma_{int} = \frac{1}{8s} \sum_{\text{spins}} 2\,\mbox{Re}[\mathcal {M}_B^+ \mathcal {M}_\psi] \, d\Phi_2,
    \label{dsigmaint}
\ea
Here my goal is to obtain the contribution to the total cross section.

Therefore, after performing integration over the phase space of the final particles in \eqref{dsigmaint},
one obtains the following formulas for the contribution to the total cross section:
\ba
    \sigma_{int}(s)
    &=
    \frac{1}{4 s^2} \mbox{Re}\biggr\{
    	\frac{\sum_{s}
    	(J^{e\bar{e}\to\gamma}_\mu)^* J^{e\bar{e}\to\psi}_\nu}
    	{s-M_\psi^2+i M_\psi\,\Gamma_\psi} \cdot
    	\sum_{s'}
    	\int d\Phi_2
    	(J_{\gamma\to \Xi^0\bar{\Xi}^0}^\mu)^* J_{\psi\to \Xi^0\bar{\Xi}^0}^\nu \biggr\},
    \label{sigmaint}
\ea
where $\sum_s$ and $\sum_{s'}$ are the summation over the spin states of the initial and final particles, respectively.
Here after using the procedure of the invariant integration method over the total volume of the final phase,
for the second term in \eqref{sigmaint} one obtains the expression in the following form:
\ba
	&\sum_{s'} \int d\Phi_2	(J_{\gamma\to \Xi^0 \bar{\Xi}^0}^\mu)^* J_{\psi\to \Xi^0 \bar{\Xi}^0}^\nu
	= \frac{1}{3} \biggl( g^{\mu\nu} - \frac{q^\mu q^\nu}{q^2} \biggr)
	\sum_{s'} \int d\Phi_2 	\br{J_{\gamma\to \Xi^0 \bar{\Xi}^0}^\alpha}^* J^{\psi \to \Xi^0 \bar{\Xi}^0}_\alpha.
\label{JJ}
\ea
In these cases, I use explicit expressions for lepton currents  $J^{e\bar{e}\to\gamma}_\mu$ from \eqref{BornAmplitude} and
$J^{e\bar{e}\to\psi}_\nu$ from \eqref{Jeepsi} and take into account the law of conservation of currents;
then, based on this, I can calculate
\ba
&&	\sum_{s}(J^{e\bar{e} \to \gamma}_\mu)^* J_{e\bar{e} \to \psi}^\mu
 = -e g_e \sum_{s} [\bar{u}(p_-) \gamma_\mu v(p_+)] [\bar{v}(p_+) \gamma^{\mu} u(p_-)] \approx  \nn \\
&&  \approx
	-e g_e \, \Tr[\hat{p_-} \gamma_\mu \hat{p_+} \gamma^\mu] \approx 4 \, e \, g_e s.
\label{JJe}
\ea
Using the expression for the two-particle phase volume of the final particles from \eqref{Phi2}, the invariant integration
method from \eqref{JJ} and the result from \eqref{JJe}, substituting all this into formula \eqref{sigmaint},
for the interference contribution to the total cross section I then obtain the expression in the form:
\ba
    \sigma_{int}(s)
    = \frac{e g_e \beta}{48 \pi s}
       \mbox{Re}\biggl\{\frac{1}{s-M_\psi^2+i M_\psi\, \Gamma_\psi}
    	\int\limits_{-1}^1 d\cos\theta_{\Xi^0} \sum_{s'}
    	(J_{\gamma\to \Xi^0 \bar{\Xi}^0}^\alpha)^* J^{\psi \to \Xi^0 \bar{\Xi}^0}_\alpha \biggr\}.
    \label{TotalCrossSectionInterference}
\ea
The subintegral expression in \eqref{TotalCrossSectionInterference} can be represented in a separate form, since it contains
the dynamics of the transformation of charmonium into a pair, $\Xi^0\bar{\Xi}^0$, that is
\ba
	S_i(s)
    = \frac{e g_e \beta}{48 \pi s}
    	\int\limits_{-1}^1 d\cos\theta_{\Xi^0} \sum_{s'}
    	(J_{\gamma \to \Xi^0\bar{\Xi}^0}^\alpha)^* J^{\psi \to \Xi^0 \bar{\Xi}^0}_\alpha.
    \label{Si}
\ea
Thus, if I take into account formula \eqref{Si} in \eqref{TotalCrossSectionInterference}, then for the
interference contribution to the total cross section I can represent $\sigma_{int}(s)$ as follows:
\ba
	\sigma_{int}(s)	=
	\mbox{Re}\biggl(\frac{S_i(s)}{s-M_\psi^2+i M_\psi\, \Gamma_\psi}\biggr),
	\label{SigmaIntViaSi}
\ea
Different mechanisms of this transformation in formulas \eqref{Si} and \eqref{SigmaIntViaSi} are denoted by $i$ in the subscript.
I know that the mass of $\psi(3770)$ exceeds the threshold for the production of a $D\bar{D}$ pair; thus, it is natural to expect
the $D$-meson loop to be the main mechanism in this reaction.
Considering this, in Fig.~\ref{fig.DD} I present a Feynman diagram, which describes quite possibly the usual OZI-permitted
mechanism via the $D$-meson loop.

However, according to this mechanism, the threshold excess is very small, i.e., the mass $\psi(3770)$ slightly exceeds the threshold
of pair production of $D$-mesons (it can be seen that $\br{M_\psi - 2M_D}/M_\psi \approx 1\%$).
For this, I suggest considering the OZI breaking mechanism due to the charmonium three-gluon annihilation into the
$\Xi^0\bar{\Xi}^0$ pair (see Fig.~\ref{fig.3G}), which gives the dominant contribution.

Thus, given the interference contribution (\ref{SigmaIntViaSi}) with the total relative phase between the Born amplitude
$\mathcal{M}_B$ and the charmonium contribution $\mathcal{M}_\psi$, one can restore the total cross section.
In the present case, I will use the procedure described in \cite{Bystritskiy} [equations~(15) and (16)].

% =======================================================================
\section{The \textit{D}-meson loop mechanism}
\label{sec.DMesonLoopMechanism}
% =======================================================================

In this process, I will also follow our previous calculations \cite{Ahmadov1,Bystritskiy,BA,Ahmadov2,Ahmadov3} with only
one systematic modification which is necessary to describe the final state of the $\Xi^0\bar{\Xi}^0$ pair.
Thus, in this section, I want to consider the contribution of the intermediate charmonium with the transition to the final
state of the $\Xi^0\bar{\Xi}^0$ hyperon pair via the $D$-meson loop.
In Fig.~\ref{fig.DD}, I can illustrate the Feynman diagram for the production of $\Xi^0\bar{\Xi}^0$ via the $D$-meson loop mechanism.
To calculate $\sigma_{int}(s)$ from \eqref{SigmaIntViaSi}, I first need to calculate the contribution of the $D$-meson loop,
that is, the quantity $S_D$ from \eqref{Si}.
Thus, according to Feynman's rules, I can write the amplitude $\mathcal {M}_D$ (according to Fig.~\ref{fig.DD}) and
from this amplitude I can extract the value $S_D$.
The amplitude $\mathcal {M}_D$ I can be written as follows:
\begin{figure}
	\centering
    \includegraphics[width=0.50\textwidth]{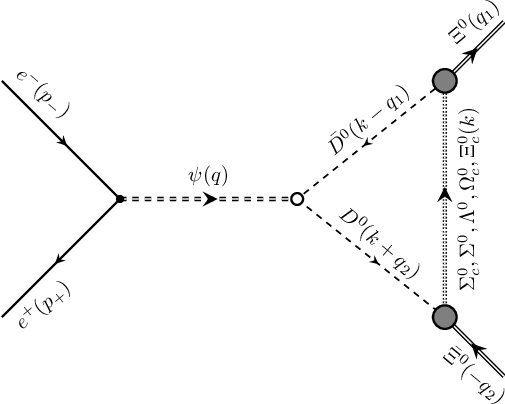}
    \caption{The Feynman diagram for the $\Xi^0\bar{\Xi}^0$ hyperon pair production in charmonium decays
    through the $D$-meson loop mechanism in the $e^+e^-$  annihilation process at the one-loop level.}
    \label{fig.DD}
\end{figure}
\ba
&&	\mathcal{M}_D =
    \frac{g_e}{16 \pi^2}
    \frac{[\bar{v}(p_+) \gamma_\mu u(p_-)]}{q^2-M_\psi^2+i M_\psi\,\Gamma_\psi}
    \cdot
    \int\frac{dk}{i \pi^2}
    \frac{[\bar{u}(q_+) \gamma_5 (\hat{k} + M_\Sigma) \gamma_5 v(q_2)] (2 k + q_2 - q_1)^\mu}{(k^2 - M_\Sigma^2)((k-q_1)^2 - M_D^2) ((k+q_2)^2 - M_D^2)}
    \times\nn\\
&& \times G_{\psi D\bar{D}}(q^2,(k+q_2)^2,(k-q_1)^2) G_{\Sigma D \Xi}(k^2,(k-q_1)^2) G_{\Sigma D \Xi}(k^2,(k+q_2)^2),
    \label{AmplitudePsiDD}
\ea
where $M_\psi$ is the mass of the $\psi (3770)$ charmonium, $M_D$ is the mass of the $D$-meson, and $M_\Sigma$ is the mass of the $\Sigma$ hyperon.
Thus, for the vertex $\psi D\bar{D}$, I use the form factor dependence in this form:
\ba
	G_{\psi D\bar{D}} (s,M_D^2,M_D^2) =
	g_{\psi D\bar{D}} \, \frac{M_\psi^2}{s} \, \frac{\log (M_\psi^2/\Lambda_D^2)}{\log (s/\Lambda_D^2)},
	\label{PsiDDFormfactor}
\ea
where the scale $\Lambda_D$ is fixed on the characteristic value of the reaction $\Lambda_D = 2 M_D$.
When comparing the expression \eqref{AmplitudePsiDD} with the general form of the amplitude from \eqref{Mpsi}, for the current
$J^{\psi\to \Xi^0\bar{\Xi}^0}_\nu (q)$ I can obtain the following expression:
\ba
&& J^{\psi\to \Xi^0 \bar{\Xi}^0}_\mu (q) = \frac{1}{16 \pi^2}
\int\frac{dk}{i \pi^2}
    \frac{[\bar{u}(q_+) \gamma_5 (\hat{k} + M_\Sigma) \gamma_5 v(q_2)] (2 k + q_2 - q_1)^\mu}{(k^2 - M_\Sigma^2)((k-q_1)^2 - M_D^2) ((k+q_2)^2 - M_D^2)}
   \times\nn\\
&& \times G_{\psi D\bar{D}}(q^2,(k+q_2)^2,(k-q_1)^2) G_{\Sigma D \Xi}(k^2,(k-q_1)^2) G_{\Sigma D \Xi}(k^2,(k+q_2)^2),
\label{JDD}
\ea
which is included in \eqref{Si}.
It is worth noting that the $D$-meson loop mechanism (Fig.~\ref{fig.DD}) with the Born amplitude contributes to the
interference of a charmonium state (see \eqref{Si}).
In this case, for $S_D$ \eqref{Si} I obtain an expression as follows:
\ba
&&	S_D(s) = \frac{\alpha g_e \beta G(s)}{48 \pi^2 s} \int\frac{dk}{i \pi^2}
    \frac{Tr D(s, k^2)}{(k^2 - M_\Sigma^2)((k-q_1)^2 - M_D^2) ((k+q_2)^2 - M_D^2)}
   \times\nn\\
&&  \times G_{\psi D\bar{D}}(q^2,(k+q_2)^2,(k-q_1)^2) G_{\Sigma D \Xi}(k^2,(k-q_1)^2) G_{\Sigma D \Xi}(k^2,(k+q_2)^2 = \nn \\
&&  \alpha_{D}(s) Z_D(s).
   \label{SDG}
\ea
From this formula, the quantities $\alpha_{D}(s)$ and $Z_D(s)$ can be written separately, which have the corresponding forms
\ba
&& \alpha_{D}(s) = \frac{\alpha g_e \beta G(s)}{48 \pi^2}, \nn \\
&& Z_D(s) = \frac{1}{s} \int\frac{dk}{i \pi^2}
    \frac{Tr D(s, k^2)}{(k^2 - M_\Sigma^2)((k-q_1)^2 - M_D^2) ((k+q_2)^2 - M_D^2)}
   \times \nn \\
&&  \times G_{\psi D\bar{D}}(q^2,(k+q_2)^2,(k-q_1)^2) G_{\Sigma D \Xi}(k^2,(k-q_1)^2) G_{\Sigma D \Xi}(k^2,(k+q_2)^2.
  \label{ZD}
\ea
where $Tr D(s, k^2)$ is the trace of the $\gamma$ matrix along the baryon line, which can be written as follows:
\ba
&&	Tr D(s, k^2) =
     Tr [(\hat {q_+}+M_\Xi) \gamma_5 (\hat {k}+M_\Sigma) \gamma_5 (\hat {q_2}-M_\Xi) (\hat {k} - M_\Xi)]
    =\nn\\
&&  = 2\left((k^2)^2 + k^2 (s - 2 (M_D^2 + M_\Xi M_\Sigma)) - s M_\Xi M_\Sigma + c_D  \right),
    \label{SpDExplicit}
\ea
where $c_D$ has the form
\ba
&c_D
    =
    M_D^4 + 2 M_\Xi M_\Sigma M_D^2 + 2 M_\Sigma M_\Xi^3 - M_\Xi^4.
    \label{cD}
\ea
In formula \eqref{SDG}, the quantities $G_{\psi D\bar{D}}$ and $G_{\Xi D \Sigma}$ for the vertices
$\psi \to D\bar{D}$ and $D \to \Xi \Sigma$ are the form factors \cite{Aubert762007,Ablikim8312022,Ablikim052024,Ablikim112024,Gong2023}.

Now I need to compute the quantity of $Z_D(s)$ from \eqref{ZD}.
For this aim I use the Cutkosky cutting rules \cite{Cutkosky} for the $D$-meson propagators in \eqref{ZD} to equivalently
replace them through the delta function
\ba
  \frac{1}{(k-q_1)^2 - M_D^2} \frac{1}{(k+q_2)^2 - M_D^2} &\longrightarrow (- 2 \pi i)^2 ~ \delta ((k-q_1)^2 - M_D^2) ~\theta (-(k-q_1)_0) \cdot \nn \\
&\cdot \delta ((k+q_2)^2 - M_D^2) ~\theta ((k+q_2)_0).
\label{delta}
\ea
Then I use these delta functions \eqref{delta} and get the imaginary part of this quantity from $Z_D(s)$:
\ba
&&	2i \, \mbox{Im} \, Z_D(s)
	=\frac{(-2\pi i)^2}{s}
    \int\frac{dk}{i \pi^2}
    \frac{Tr D(s, k^2)}{k^2 - M_\Xi^2}
    G_{\psi D\bar{D}}(s,(k+q_2)^2,(k-q_1)^2)
    \times\nn\\
&&   \times
    G_{\Sigma D \Xi}(k^2,(k-q_1)^2)~G_{\Sigma D \Xi}(k^2,(k+q_2)^2)
    \delta ((k+q_2)^2 - M_D^2)~\delta ((k-q_1)^2 - M_D^2)
    \times\nn\\
&&    \times
    \theta ((k+q_2)_0)~\theta (-(k-q_1)_0).
    \label{ImZD0}
\ea
Note that by replacing these two delta functions in (\ref{ImZD0}) and implementing cyclic integrations to
significantly simplify the evaluation of $\mbox{Im} \, Z_D$, I obtain the final expression for the imaginary
part of this value in this form:
\ba
&&	\mbox{Im} \, Z_D\br{s}=
	-\frac{2\pi}{s^{3/2}}
	G_{\psi D\bar{D}}(s,M_D^2,M_D^2)
	\int\limits_{C_k^{(1)}}^1 \frac{dC_k}{\sqrt{D_1}}
    \sum_{i=1,2}
    \frac{k_{(i)}^2} {k_{(i)}^2 + M_\Xi^2}~
	\times\nn\\
&& \quad\times
    Tr D(s,-k_{(i)}^2)~G_{\Sigma D \Xi}^2(-k_{(i)}^2,M_D^2).
        \label{ImZD}
\ea
Here $s > 4M_D^2$, whereas integration over the cosine of the polar angle $C_k = \cos\theta_k$ is calculated numerically.
To do this, I can define the values of $k_{(i)}$, $D_1$, and $C_k^{(1)}$ as
\ba
	k_{(1,2)} = \frac{1}{2} (\sqrt{s} \, \beta \, C_k \pm \sqrt{ D_1 }),
	\,\,\,
	D_1 = s \, \beta^2 \, C_k^2 - 4 (M_D^2 - M_{\Xi}^2),
    \,\,\,
    C_k^{(1,2)} = \pm \frac{2}{\beta} \sqrt{\frac{M_D^2 - M_{\Xi}^2}{s}}.
\ea
Now I must consider the explicit expression of the form factors to perform the calculations of $\mbox{Im}\, Z_D$ from (\ref{ImZD}).
It should be noted that for the vertex $\psi\to D\bar{D}$ I only need the dependence on the charmonium virtuality $q^2=s$.
In this case, I assume that the legs of the $D$ meson are on-mass-shell, and then using expression (\ref{ImZD}),
I calculate the imaginary part of $Z_D$.

I would like to note that the details of the calculation of $Z_D\br{s}$ are in \cite{Ahmadov1,Bystritskiy}; however,
in the present work I technically compute the imaginary part of this quantity and then by apply the dispersion relation
with one subtraction at $q^2=0$, and  after that one can restore its real part.
Also, it is worth noting that the subtraction constant here also vanishes, since the $\Xi$ hyperon (which is the $uss$ quark state)
has no open charm and, therefore, at $q^2=0$ the vertex $\psi \to \Xi^0 \bar{\Xi}^0$ is zero.

The normalizations of the function $G_{\psi D\bar{D}}(s,M_D^2,M_D^2)$ to the decay of the charmonium $\psi(3770)$ into the
final state $D\bar{D}$ must be fixed.
The decay width $\psi \to D \bar{D}$ is used to fix these functions:
\vspace*{-0.5cm}
\ba
g_{\psi D\bar{D}} \equiv G_{\psi D\bar{D}}(M_\psi^2, M_D^2, M_D^2).
\ea
To have a vertex $\psi \to D\bar{D}$ with the only dependence on charmonium virtuality $q^2 = s$, it is necessary
to cut the diagram by $D$-meson propagators, since the $D$-meson legs are on the mass shell.

To determine the value of the constant $g_{\psi D\bar{D}}$, I need to know the decay widths of the charmonium $\psi \to D \bar{D}$.
The standard formula for calculating the total decay width can be written as follows:
\ba
\Gamma_{\psi D\bar{D}} = \frac{g_{\psi D\bar{D}}^2 M_{\psi}\beta_{D}^3}{48 \pi}.
\ea
Now I can calculate the value of the constant $g_{\psi D\bar{D}}$.
For this one can use the experimental value of the decay width of the charmonium $\Gamma_{\psi D\bar{D}} = 25 \,MeV$ \cite{PDG} and
then compute the constants $g_{\psi D\bar{D}}$, and accordingly obtain the value:
\ba
g_{\psi D\bar{D}} = 4 \sqrt{\frac{3 \pi \Gamma_{\psi D\bar{D}}}{M_{\psi}\beta_{D}^3}} \approx 18.4,
\ea
where $\beta_D = \sqrt{1 - 4M_D^2/M_\psi^2}$ is the $D$-meson velocity in the $\psi \to D\bar{D}$ decay.

I would like to point out that in \cite{Ahmadov1,Bystritskiy,BA} we used the $\Lambda D P$-vertex
based on the results of \cite{Reinders,Navarra}.
In \cite{Ahmadov2,Ahmadov3}, I used a constant for the $\Xi D \Sigma$-vertex, which is based on the results of \cite{Reinders,Choe}.
However, in the present paper I use for the $\Sigma D \Xi $ vertex another form of constant according to the results of \cite{Reinders,Choe}.

Now I need to consider the function $G_{\Sigma D \Xi} (k^2,p^2)$ from (\ref{ZD}).
I know that the only dependence left after the application of the Cutkosky rule is the off-mass-shellness of the $\Sigma^0_c$ hyperon
in the scattering regime, since here $k^2 = -k_{(i)}^2 < 0$.
The dependence over the virtuality of the $D$ meson has disappeared, $D$ mesons are on mass shell.
However, I still need to take into account the remnants of this dependence.
I use this form for the $\Sigma D \Xi$ vertex with the off-mass-shell $D$ meson which was established in \cite{Navarra,Reinders,Choe}:
\ba
 g_D(p^2) = \frac{2 M_D^2 f_D}{m_u + m_c}\frac{g_{\Sigma D \Xi}}{p^2 - M_D^2},
\ea
where $f_D \approx 180~\MeV$.\\
Thus in the calculation in Eq.\eqref{ImZD} I use the following expression:
\ba
G_{\Xi D \Sigma}(-k_{(i)}^2, M_D^2) = \frac{f_D \, g_{\Sigma D \Xi}}{m_u + m_c}.
\label{GXDS}
\ea
It should be noted that Eq.\eqref{GXDS} represents a vertex function with a $D$ meson on-mass-shell, since in Eq.\eqref{ImZD}
I estimate imaginary part of \mbox{$Z_D$} (see Eq.\eqref{delta}).

The constant $g_{\Sigma D \Xi}$ was estimated in \cite{Choe} in the scattering regime, i.e. for $p^2 < 0$ and
\ba
	g_{\Sigma D \Xi} \approx g_{K \Sigma \Xi} = -7.02,
	\label{gLambdaDXi}
\ea
which do not take the effects of the $\Sigma^0_c$-hyperon off-mass-shellness.
I expect that these effects are not very important.

For the mass of $u$ and $c$ quarks, I apply the following values: $m_u\approx 280~\MeV$ and $m_c=1.27~\GeV$ \cite{PDG}.

% ========================================================================================================================
\section{The three-gluon mechanism}
\label{sec.ThreeGluonMechanism}
% ========================================================================================================================

\begin{figure}
	\centering
    \includegraphics[width=0.50\textwidth]{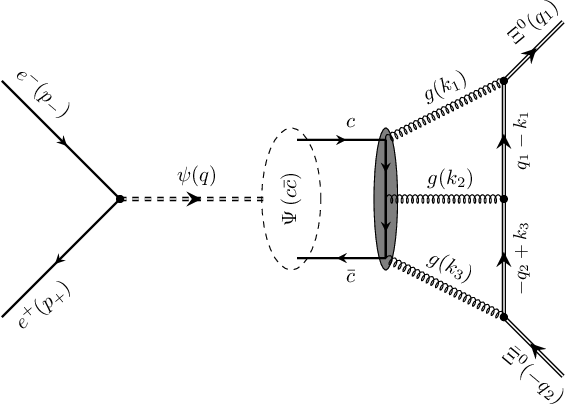}
    \caption{The Feynman diagram of the $\Xi^0\bar{\Xi}^0$ hyperon pair production in charmonium decays
    through the three-gluon mechanism in the $e^+e^-$ annihilation process.}
    \label{fig.3G}
\end{figure}

In here, I study how the intermediate charmonium contributes to the production of the $\Xi^0 \bar{\Xi}^0$
pair through the three-gluon annihilation mechanism.
In this case I calculate the quantity of $S_{3g}$ from \eqref{Si} to determine the contribution of $S_{3g}$ to the cross section \eqref{SigmaIntViaSi}.
The Feynman diagram for the production of a $\Xi^0 \bar{\Xi}^0$ hyperon pair via the three-gluon mechanism in the $e^+e^-$
annihilation is presented in Fig.~\ref{fig.3G}.
It is necessary to emphasize that a reaction similar to the present, i.e., the production of a pair of hadrons
in the process of annihilation of $e^+e^-$ through the three-gluon mechanism, was considered in
\cite{Ahmadov1,Bystritskiy}; it was revised and some typos and minor errors were corrected.
In this paper, I apply this mechanism to this process via the three-gluon exchange.

Therefore, I present the final formulas for the contribution to the quantity of $S_{3g}$ from \eqref{SigmaIntViaSi}
in the interference of the charmonium state with the Born amplitude [see (\ref{Si})] according to the relevant Feynman diagram
(Fig.~\ref{fig.3G}), [which coincide with Eqs. (16) and (17) from \cite{Ahmadov1}]:
\ba
    S_{3g}(s) = \alpha_{3g} (s) \, Z_{3g}(s),
    \label{S3g}
\ea
where
\ba
&&    \alpha_{3g} (s) = \frac{\alpha\, \alpha_s^3}{2^3 \, 3} g_e \, g_{col} \, \phi \, \beta \, G (s) \, G_\psi(s),
    \label{alpha3g}
    \\
&&    Z_{3g} (s)= \frac{4}{\pi^5 s}\int \frac{dk_1}{k_1^2}\frac{dk_2}{k_2^2}\frac{dk_3}{k_3^2}
    \frac{Tr3g ~ \delta (q-k_1-k_2-k_3)}{((q_1 - k_1)^2 - M_\Xi^2) ((q_2 - k_3)^2 - M_\Xi^2)},
    \label{Z3g}
\ea
where $g_{col}$ is the color factor, $g_{col} (1/4) = \left<\Xi| d^{ijk} ~ T^i T^j T^k |\Xi \right> = 15/2$.
The $Tr3g$ is the product of traces over the $\Xi$-hyperon and $c$-quark lines,
\ba
&&	Tr3g = \Tr [\hat{Q}_{\alpha\beta\gamma} (\hat {k}_c + m_c) \gamma^\mu (\hat {k}_{\bar{c}} - m_c)] \cdot
    \Tr\left[(\hat {q}_1 +M_\Xi) \gamma^\alpha (\hat{q}_1 - \hat {k}_1 + M_\Xi) \gamma^\beta \right.
	\times     \nn \\
&& \qquad\times \left. (-\hat {q}_2 + \hat{k}_3 + M_\Xi) \gamma^\gamma (\hat {q}_2 -M_\Xi) \gamma_\mu \right],
   	\nn
\ea
where
\ba
&& \hat{Q}_{\alpha\beta\gamma}  =
    \frac{\gamma_\gamma (-\hat {k}_{\bar{c}} + \hat {k}_3 + m_c) \gamma_\beta (\hat {k}_c - \hat {k}_1 + m_c) \gamma_\alpha}
    {((k_{\bar{c}}-k_3)^2 - m_c^2) ((k_c-k_1)^2 - m_c^2)} +[\mbox{gluon permutations}].
\label{DefinitionHatQ}
\ea
I would like to note that the quantity $\phi$ in (\ref{alpha3g}) is related to the charmonium wave function and can be written in the following form:
\ba
	\phi = \frac{|\psi (\bf {r}=\bf{0})|}{M_\psi^{3/2}} = \frac{\alpha_s^{3/2}}{3\sqrt{3\pi}},
	\label{eq.phi}
\ea
where $\alpha_s$ is the QCD coupling constant.
Here the quantity $\phi$ is obtained from the decay rate of $\psi\to 3\,g$ on the mass shell.
On the charmonium scale (for $s \sim M_c^2$) this three-gluon mechanism depends on its
value to a rather high degree, and therefore, this mechanism, according to Eqs.\eqref{alpha3g} and
\eqref{eq.phi}, is very sensitive to this value of $\phi$.
For this process, when performing computations I use the value $\alpha_s(M_c) = 0.28$ which is expected
at the QCD evolution of $\alpha_s$ from the $b$-quark scale to the $c$-quark scale, i.e., to $s \sim M_c^2$.

It is necessary to note that one of the most important corrections concerns the final state of $\Xi^0 \bar {\Xi}^0$.
During the execution of the process, after the decays of $\psi(3770)$, three gluons are obtained, then three more pairs
of the quark-antiquark are formed and in the final state $\Xi^0 \bar {\Xi}^0$ pairs are formed.
To completely implement this mechanism, it is necessary to reproduce the absolute value of the cross section.
One of the goals here of this mechanism is the transition of three gluons (i.e., with a total angular momentum of $1$)
into a final $\Xi^0 \bar {\Xi}^0$ pair.

Based on the computing method, I can assume that this mechanism has much in common in the timelike region with the
production of $p \bar {p}$, $\Lambda \bar {\Lambda}$, $\Sigma^0 \bar {\Sigma}^0$ and $\Sigma^+ \bar {\Sigma}^-$ pairs
from a photon \cite{Bystritskiy,BA,Ahmadov2,Ahmadov3}.
In formula (\ref{alpha3g}) I take the factor $G_\psi(s)$, describing the mechanism of the transition of three
gluons into the final pair $\Xi^0 \bar{\Xi}^0$, as a form factor.
Thus, similar to (\ref{Formfactor}), in (\ref{alpha3g}) I can apply an additional form factor with a different value
of the parameter $C_\psi$, that is
\ba
	|G_\psi(s)| = \frac{C_\psi}{s^2 \log^2 (s/\Lambda_{\text{QCD}}^2)}.
	\label{FormfactorPsi}
\ea
In this process, for the parameter $C_\psi$ I use the same value that was used in the case of the production
of the $p \bar {p}$ and the $\Lambda \bar {\Lambda}$ \cite{Bystritskiy,BA} pairs:
\ba
	C_\psi = (45 \pm 9)~\GeV^4.
	\label{eq.CPsi}
\ea
Now it is necessary to restore the real part of $Z_D$ from (\ref{ZD}) and $Z_{3g}$ from (\ref{Z3g}).
By applying the technique of dispersion relations, which is described in detail in \cite{Ahmadov1},
I further obtain an expression for the real part of $Z_D$ and $Z_{3g}$ in the form
\ba
	\mbox{Re} \, Z_i(\beta)	=
	\frac{1}{\pi} \left\{
		\mbox{Im} \, Z_i(\beta) \log \left|\frac{1-\beta^2}{\beta_{\text{min}}^2 - \beta^2}\right|
		+ \int\limits_{\beta_{\text{min}}}^1 \frac{2 \beta_1 d\beta_1}{\beta_1^2 - \beta^2}
		[\mbox{Im} \, Z_i(\beta_1) - \mbox{Im} \, Z_i(\beta)]\right\}.
	\label{DispersionRelation}
\ea
When calculating the imaginary part, I make sure that the imaginary part $\mbox{Im}\, Z_D(\beta)$ of the contribution
of the $D$- meson loop from (\ref{ImZD}) above the threshold ($s > 4M_D^2$) is not zero.
So, this means that the integration over the lower limit in \eqref{DispersionRelation} is equal to
$\beta_{\text{min}} = \sqrt{1 - M_{\Xi}^2/M_D^2}$.
However, I observe that when the value is $s_{\text{min}} = 4M_{\Xi}^2$, the imaginary part $\mbox{Im} \, Z_{3g}(\beta)$
for the three-gluon contribution matches the reaction threshold, so I can treat the lower limit of integration as $\beta_{\text{min}} = 0$.

% =================================================================================================
\section{The Numerical results}
\label{sec.Numerical}
% =================================================================================================

Now I want to describe the results obtained in the present work.
I have investigated in detail the characteristics of the distribution of the total cross section from the center-of-mass
energy of the production of $\Xi^0 \bar {\Xi}^0$ in electron-positron annihilations at BESIII energies \cite{Ablikim112024}.
In particular, according to the calculations carried out in this work, I have plotted the quantities of $Z_D (s)$ and $Z_{3g} (s)$
depending on the total energy $\sqrt {s}$ in the range starting from the reaction threshold $s=4M_{\Xi}^2$ to 4.95 GeV,
since these quantities give the corresponding ($D$-meson loop and three gluons) contributions.
Of course, I performed a comparison of all the obtained theoretical results with the experimental data of BESIII.
Based on formula \eqref{TotalCrossSectionBorn}, one can calculate the total cross section for the production of
$\Xi^0 \bar{\Xi}^0$ pairs in their annihilation process as a function of the collider center-of-mass energy in the range from 3.51 GeV to 4.95 GeV.  \\
In Fig.~\ref{WideRange}, the dependence of the total cross section on the center-of-mass energy $\sqrt s$ of the process
$e^+ e^- \to \Xi^0 \bar{\Xi}^0$ in the Born approximation is presented.

I have plotted this graph based on the theoretical results for the total cross section in the Born approximation, and
here my theoretical result is compared with the experimental BESIII data.
It is seen that the result obtained by me agrees well with the experimental points.
It is necessary also to note that the maximum value of the total cross-section is obtained at an energy of $\sqrt s$ = 3.51 GeV;
then, with increasing energy, the total cross-section smoothly decreases.

Note that the propagator considered in my process between the vertices of $D^0 \bar{D^0}$ contains five types of hyperons:
$\Sigma^0_c, \,\Sigma^0, \,\Lambda^0, \,\Omega^0_c, \,\Xi^0_c$, i.e., there are different options in this reaction.
Of course, the hyperons formed in the final state can be obtained from different combinations of quarks.
In calculations I used the average mass of five hyperons for the hyperon mass in the propagator
(between the vertices of $D^0 \bar{D^0}$).

\begin{figure}
	\vspace{5mm}
	\centering
    \includegraphics[width=0.60\textwidth]{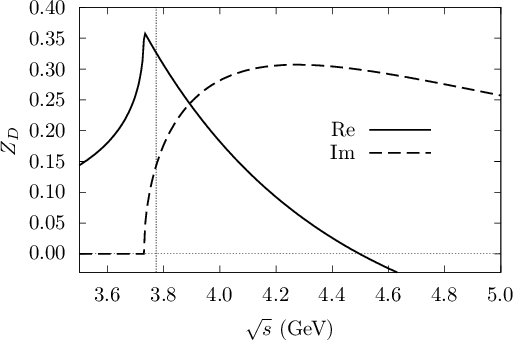}
    \caption{
    	The quantity $Z_D\br{s}$ from (\ref{ZD}) depending on the center-mass-energy $\sqrt{s}$
    starting from the threshold $\sqrt{s} = 2M_\Xi^0$.
    The position of $\psi(3770)$ is indicated by a vertical dashed line.}
    \label{fig.ZD}
\end{figure}

\begin{figure}
	\vspace{5mm}
	\centering
    \includegraphics[width=0.60\textwidth]{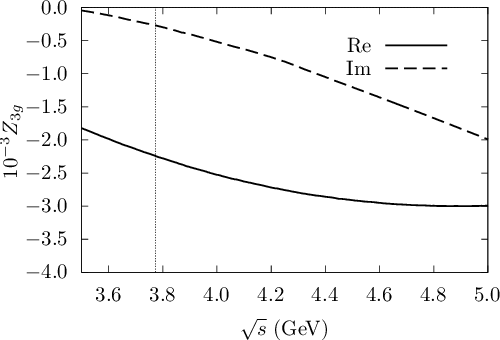}
    \caption{
    	The quantity $Z_{3g}\br{s}$ from (\ref{Z3g}) depending on the center-mass-energy $\sqrt{s}$
    starting from the threshold $\sqrt{s} = 2M_\Xi^0$.
    The position of $\psi(3770)$ is indicated by a vertical dashed line.}
    \label{fig.Z3g}
\end{figure}

Figure~\ref{fig.ZD} shows the dependence of the quantity of $Z_D (s)$ on the total energy $\sqrt{s}$ in the range from the
reaction threshold $\sqrt{s}= 2M_\Xi^0$ to $4.95~$GeV.
The form and numerical values of the real and imaginary parts of this quantity are almost the same as in the cases
of the production of the proton-antiproton (see Fig.~7(a) in \cite{Bystritskiy}),
$\Lambda \bar{\Lambda}$ [Fig.~6 in \cite{BA}], $\Sigma^0 \bar{\Sigma}^0$
[Fig.~6 in \cite{Ahmadov2}] and $\Sigma^+ \bar{\Sigma}^-$ [Fig.~6 in \cite{Ahmadov3}].
Figure~\ref{fig.Z3g} presents the dependence of the real and imaginary parts of $Z_{3g} (s)$ on the total energy
$\sqrt{s}$ in the range from the reaction threshold $\sqrt{s}= 2M_\Xi^0$ to $4.95~$GeV.

It can be seen from the graph that the general behavior of the $Z_{3g} (s)$ curves in $\sqrt{s}$ is very different from the
curves that were obtained in the case of the production of proton-antiprotons [Fig.~7 (B) \cite{Bystritskiy}],
$\Lambda \bar{\Lambda}$ [Fig.~7 \cite{BA}], and $\Sigma^0 \bar{\Sigma}^0$ [Fig.~7 in \cite{Ahmadov2}],
but it shows a similar general behavior of the curves as in the case of the $\Sigma^+ \bar{\Sigma}^-$ final state
(Fig.~7 in \cite{Ahmadov3}).
Nevertheless, it can be seen from the graph that the characteristic large negative quantity of $Z_{3g}(s)$ still remains the same,
since it gives a large relative phase with respect to the Born contribution in the amplitude.
Taking into account the above, one can make sure that the quantities $Z_D (s)$ from \eqref{ZD} and $Z_{3g}\br{s}$ from \eqref{Z3g}
are considered to be the main parts of the total cross section that give the corresponding contributions ($D$-meson loop and trigluons).
In Figs.\ref{fig.ZD} and \ref{fig.Z3g}, the resonance position $\psi(3770)$ is marked with a vertical dotted line.
\begin{figure}
	\centering
    \includegraphics[width=0.55\textwidth]{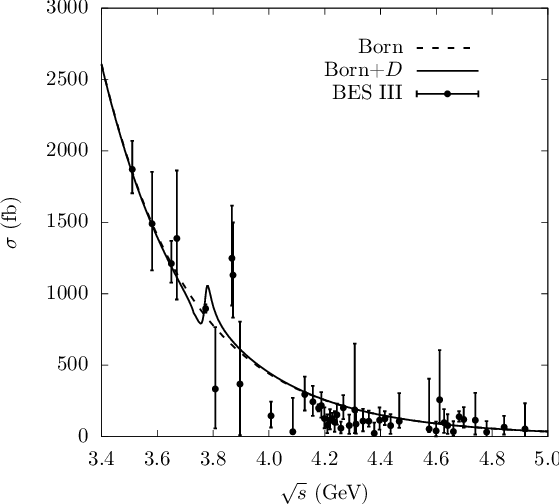}
    \caption{Total cross section distributions of the process $e^+ e^- \to \Xi^0 \bar{\Xi}^0$ as a function
    of the center-of-mass energy $\sqrt {s}$ with the inclusion of the $D$-meson loop contributing.
    The Experimental data are from BESIII Collaboration \cite{Ablikim112024}.}
    \label{fig.NarrowDLoop}
\end{figure}
\begin{figure}
	\centering
    \includegraphics[width=0.50\textwidth]{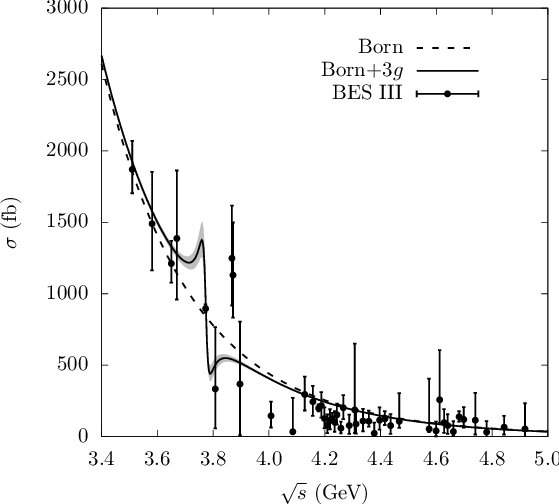}
    \caption{The total cross section distributions of the process $e^+ e^- \to \Xi^0 \bar{\Xi}^0$ as a function
    of the center-of-mass energy $\sqrt {s}$ with the inclusion of three-gluons contributing.
    The Experimental data are from BESIII Collaboration \cite{Ablikim112024}.}
    \label{fig.Narrow3g}
\end{figure}
\begin{figure}
	\centering
    \includegraphics[width=0.50\textwidth]{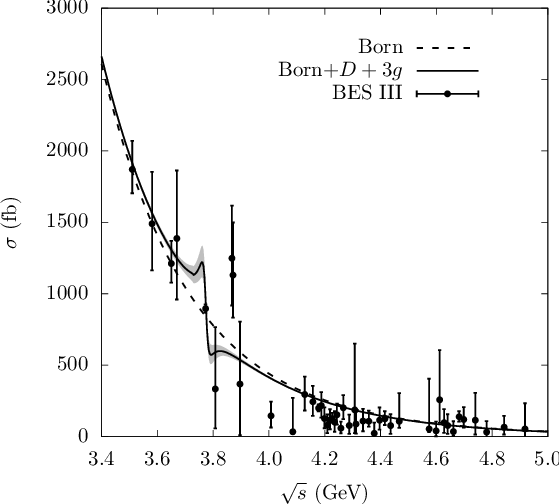}
    \caption{The total cross section distributions of the process $e^+ e^-\to \Xi^0 \bar{\Xi}^0$ as a function
    of the center-of-mass energy $\sqrt{s}$ including two mechanisms
        ($D$-meson loop and three gluon). The Experimental data are from BESIII Collaboration \cite{Ablikim112024}.}
    \label{fig.CSNarrow2}
\end{figure}
The theoretical results for the contribution of the $D$-meson loop to the Born cross section are presented in Fig.\ref{fig.NarrowDLoop}.
Similarly, in Fig.\ref{fig.Narrow3g} I show the result of the three-gluon contribution to the Born cross section.
My theoretical results, which I show in Figs.\ref{fig.NarrowDLoop} and \ref{fig.Narrow3g} for both of these contributions,
are compared with the experimental data of the BESIII collaboration \cite{Ablikim112024}.

From Fig.~\ref{fig.NarrowDLoop} I see that the experimental point at $\sqrt{s} = 3770$ GeV, i.e., the charmonium resonance point $\psi(3770)$,
almost coincides with the theoretical results I have obtained.
The cross-section for the contribution of the three-gluon mechanism, which is presented in Fig.~\ref{fig.Narrow3g},
shows that the experimental result at the point $\sqrt{s} = 3770$ GeV, i.e., the charmonium resonance point $\psi(3770)$,
completely coincides with my theoretical result.
Finally, I have studied the total cross section taking into account both contributions, i.e., the $D$-meson loop and the three-gluon mechanism,
which I show in Fig.~\ref{fig.CSNarrow2}, and from this it can be seen that the experimental result at the charmonium resonance point
$\psi(3770)$ completely coincides with my theoretical result.

The theoretical results of the total cross-section obtained by us in the energy range $\sqrt{s}$ = 3.5 - 5.0 GeV by
taking into account the contributions of the $D$-meson loop and three-gluon mechanisms in the Born cross section
and comparisons with the BESIII \cite{Ablikim112024} data are shown in Fig.~\ref{fig.CSNarrow2}.
I want to note that in this graph, the peak is near the resonance $\psi(3770)$ of charmonium and is clearly visible.
According to my model, I mainly consider the $\psi(3770)$ vicinity.
In addition, for the reaction $e^+e^- \to \Xi^0\bar{\Xi}^0$, I study the vicinity with new charmoniumlike states in the same energy region,
for example, $\psi(4040)$, $\psi(4160)$, $Y(4230)$, $Y(4360)$, $\psi(4415)$, and $Y(4660)$.
It is necessary to note that I have the $\psi(3770)$ charmonium in the intermediate state, but besides that, I calculate and investigate
the total cross section of this process taking into account new charmoniumlike states in the intermediate state.

Recall that in the process I am looking at, $e^+e^-\to \Xi^0 \bar{\Xi}^0$, I do not make any additional parameter fitting; instead,
I keep all the necessary parameters of my model the same as in the process $e^+e^-\to p\bar{p}$ \cite{Bystritskiy}.

\begin{figure}
	\centering
    \includegraphics[width=0.70\textwidth]{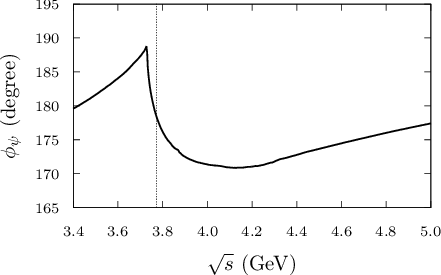}
    \caption{The dependence of the relative phase of the charmonium $\psi(3770)$ contribution from the total invariant energy $\sqrt{s}$.}
    \label{fig.PhiNarrow}
\end{figure}

Now I can determine the total relative phase $\phi_\psi$ of the charmonium contribution $\mathcal{M}_\psi$ to the amplitude relative
to the Born contribution $\mathcal{M}_B$ without taking into account the Breit-Wigner factor, i.e.,
\ba
    S_D (s) + S_{3g} (s) = |S\br{s}| e^{i \phi_\psi},
\ea
where $S_D(s)$ was defined in \eqref{SDG} and $S_{3g}(s)$ is from \eqref{S3g}.
The dependence of the total relative phase $\phi_\psi$ on the energy of the center of mass $\sqrt{s}$ is plotted in Fig.~\ref{fig.PhiNarrow}.
Here I also want to note that in this figure the resonance position of the $\psi(3770)$ charmonium is also marked with a vertical dashed line.
As can be seen from Fig.~\ref{fig.PhiNarrow}, at the point $\psi(3770)$ of charmonium, the relative phase and the corresponding
total cross section \eqref{TotalCrossSection} have the following value:
\ba
    \sigma_\psi = 1005.75~\mbox{fb},
    \qquad
    \phi_\psi = 178.45^{\circ}.
\ea
It can be concluded that for the decay of charmonium into two baryons to the finite state, such a property is common.
This has been also shown in \cite{Ahmadov1,Bystritskiy,BA}, for decay of the charmonium $\psi(3770)$ into a pair $p\bar{p}$ and a pair
$\Lambda\bar{\Lambda}$, and the $\chi_{c2}(1P)(3556)$ charmonium in \cite{Kuraev}.

In this work, for the $e^+e^- \to \Xi^0\bar{\Xi}^0$ annihilation process, in addition to the contribution
of the $\psi(3770)$ charmonium in the intermediate state, I am also exploring the contributions from charmoniumlike states,
$\psi(4040)$, \,\,$\psi(4160)$, \,\,$Y(4230)$,\,\,$Y(4360)$,\,\,$\psi(4415)$, and $Y(4660)$.

\begin{figure}
	\centering
    \subfigure[]{\includegraphics[width=0.38\textwidth]{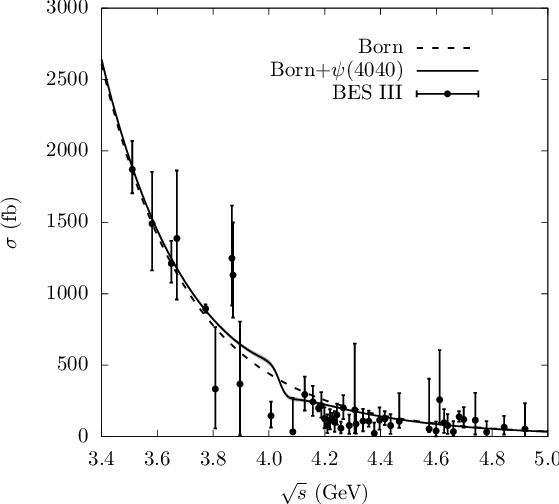}\label{psi4040}}
	\hspace{0.05\textwidth}
    \subfigure[]{\includegraphics[width=0.38\textwidth]{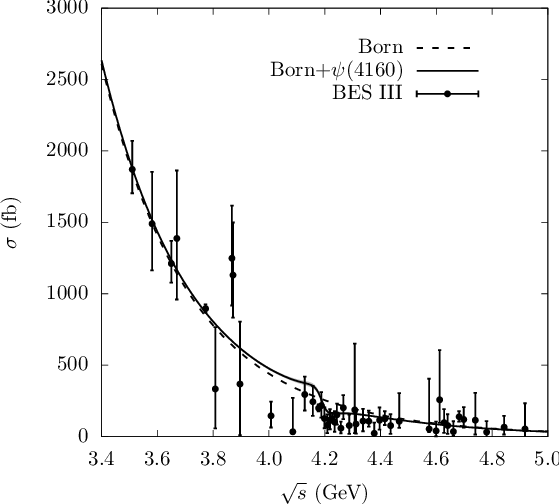}\label{psi4160}}
    \vspace{0.5cm}
    \centering
    \subfigure[]{\includegraphics[width=0.38\textwidth]{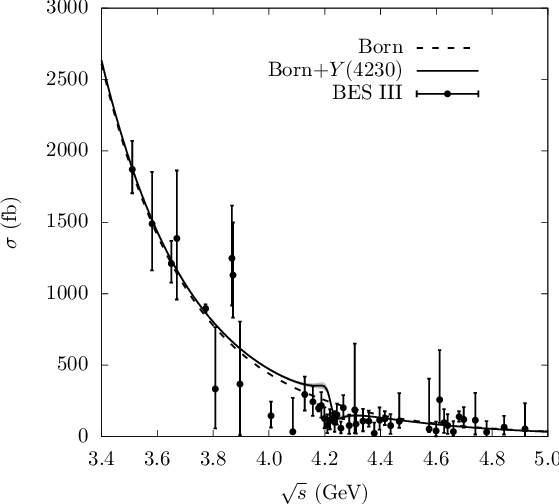}\label{Y4230}}
	\hspace{0.05\textwidth}
    \subfigure[]{\includegraphics[width=0.38\textwidth]{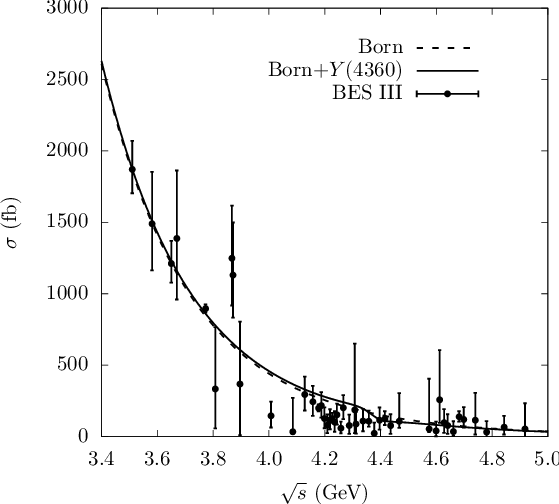}\label{Y4360}}
    \vspace{0.5cm}
    \centering
    \subfigure[]{\includegraphics[width=0.38\textwidth]{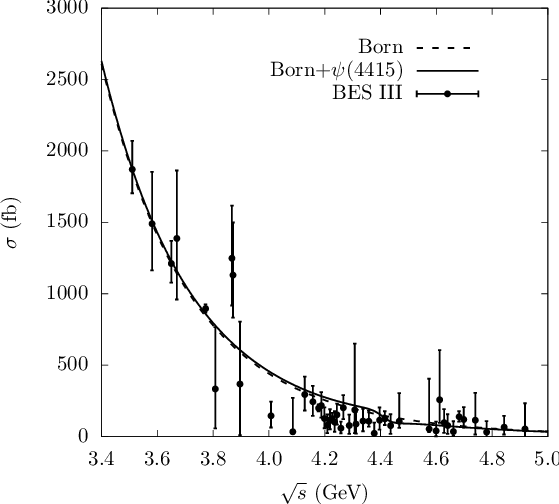}\label{psi4415}}
	\hspace{0.05\textwidth}
    \subfigure[]{\includegraphics[width=0.38\textwidth]{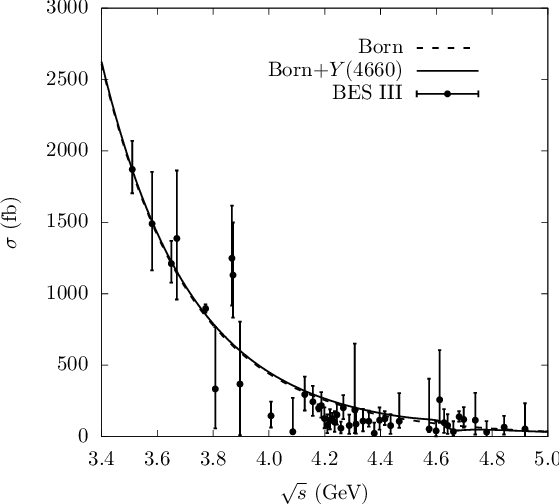}\label{Y4660}}
    \caption{The total cross section distributions of the process $e^+ e^-\to \Xi^0 \bar{\Xi}^0$ as a function of the
    center-of-mass energy $\sqrt{s}$ for the contribution of
    charmonium (-like) states, $\psi(4040)$, \,\,$\psi(4160)$, \,\,$Y(4230)$,\,\,$Y(4360)$,\,\,$\psi(4415)$, and $Y(4660)$.
    Experimental data are from BESIII Collaboration \cite{Ablikim112024}.}
    \label{CharmoniumStates}
\end{figure}

For the process $e^+e^- \to \Xi^0\bar{\Xi}^0$ I plotted the total cross section as a function of the total energy
of the collider in the region of $\sqrt{s}$ = 3.51 - 4.95 GeV based on the results of the contributions of charmoniumlike states,
i.e., respectively for $\psi(4040)$, \,\,$\psi(4160)$, \,\,$Y(4230)$,\,\,$ Y(4360)$,\,\,$\psi(4415)$ and $Y(4660)$,
and I present them in Fig.~\ref{CharmoniumStates}.
I am pleased to note that at the point for each resonance of the charmonia $\psi(4040)$, \,\,$\psi(4160)$, \,\,$Y(4230)$,\,\,
$Y(4360)$,\,\,$\psi(4415)$, and $Y(4660)$, the experimental result agrees well with my theoretical result,
which can be seen in Fig.~\ref{CharmoniumStates}.

In Table I one can see the corresponding comparison of my numerical results for each contribution (Born, $D$-loops, 3$g$)
at several selected energies, namely, from 3.765 \mbox{GeV} to 3.900 \mbox{GeV}, according to Fig.2, Fig.4, and Fig.5.
\begin{table}
\large{ \begin{tabular}{|c|c|c|c|}
  \hline
    $\sqrt {s},\, GeV$ & $\sigma ^{Born} (fb)$   & $\sigma^{D-loops} (fb)$   & $\sigma^{3g} (fb)$  \\
 \hline
  $3.765$  &  $861.346$  &  $848.505$  &  $1318.978$   \\
  \hline
  $3.778$  &  $829.653$  &  $1055.143$  &  $640.534$    \\
  \hline
  $3.785$  &  $814.276$  &  $1036.944$  &  $463.426$    \\
  \hline
  $3.791$  &  $799.205$  &  $982.893$   &  $440.812$    \\
  \hline
  $3.797$  &  $784.432$  &  $931.619$  &  $462.536$    \\
  \hline
  $3.801$  &  $777.156$  &  $909.362$  &  $476.151$    \\
  \hline
  $3.810$  &  $755.757$  &  $854.313$  &  $510.808$    \\
  \hline
  $3.820$  &  $734.989$  &  $811.575$  &  $532.801$    \\
  \hline
  $3.829$  &  $714.832$  &  $776.372$  &  $544.726$     \\
  \hline
  $3.839$  &  $695.266$  &  $746.006$  &  $549.534$     \\
  \hline
  $3.842$  &  $688.873$  &  $736.674$  &  $550.026$     \\
  \hline
  $3.852$  &  $670.068$  &  $710.501$  &  $548.820$     \\
  \hline
  $3.862$  &  $651.814$  &  $686.515$  &  $544.590$     \\
  \hline
  $3.871$  &  $634.093$  &  $664.227$  &  $538.509$    \\
  \hline
  $3.878$  &  $622.566$  &  $650.145$  &  $533.642$    \\
  \hline
  $3.884$  &  $611.265$  &  $636.600$  &  $528.079$    \\
  \hline
  $3.891$  &  $600.183$  &  $623.536$  &  $522.281$    \\
  \hline
  $3.900$  &  $583.964$  &  $604.744$  &  $513.051$    \\
  \hline
  \end{tabular}
  \caption{The numerical results for each contribution (Born, $D$-loops, 3$g$) at several selected energies.}}
 \label{tab}
\end{table}
\newpage
% =======================================================================
\section{Discussion and Conclusion}
\label{sec.Conclusion}
% =======================================================================

I have examined the production of $\Xi^0 \bar{\Xi}^0$ pairs in the process of $e^+e^-$ annihilation
in the vicinity of the charmonium resonance $\psi(3770)$ at the center-of-mass energy from 3.51 to 4.95 GeV collected
with the BESIII detector at the BEPCII collider and corresponding to an integrated luminosity of 30 $\rm{fb}^{-1}$.
I want to note that we have also studied this process near other charmonia, i.e., with new charmoniumlike states,
for example, $\psi(4040)$, $\psi(4160)$, $Y(4230)$, $Y(4360)$, $\psi(4415)$, and $Y(4660)$, in the same energy region.
I studied the production of $\Xi^0 \bar{\Xi}^0$ pairs in the process of $e^+ e^-$ annihilation at the
first stage within the framework of QED in the Born approximation.
In addition to the Born mechanism, I also studied the total cross sections taking into account two more contributions
associated with the intermediate state of the $\psi(3770)$ charmonium and other new charmoniumlike states, for example,
$\psi(4040)$, $\psi(4160)$, $Y(4230)$, $Y(4360)$, $\psi(4415)$, and $Y(4660)$.
One of these contributions is a $D$-meson loop and the second contribution is the three-gluon mechanism.
I compared the theoretical results obtained with the experimental data of BESIII \cite{Ablikim112024}.
It should be noted that their total sum gives a rather good agreement with the experimental point at $\sqrt{s} = M_{\psi,\,Y}$.
I have already mentioned that the process $e^+ + e^-\to\Xi^0 +\bar{\Xi}^0$ can also be implemented by the
vector charmonium state $\psi(3770)$ as well as by other charmoniumlike states.
The photon, $\psi$, and other charmoniumlike states are vector mesons; thus, the structures of the
cross section distribution were similar.

Indeed, in this $e^+ + e^-\to \Xi^0 +\bar{\Xi}^0$ process both mechanisms make a significant contribution,
and they explain the main part of the final result.
In turn, one of the important results is that the curve I obtained reproduces the tendency of the experimental points
in the left and right shoulders relative to the central point.
When performing computations in this process, I did not use any procedures for fitting.
For this purpose, all necessary parameters were used fixed for the $p \bar {p}$ production channel in \cite{Bystritskiy}.
I aimed to implement a fine scan with small steps of the energy area around the charmonium resonance $\psi(3770)$.

Thus, given the above, I can reasonably conclude that in our paper, as in other works, in the charmonium decay the vertex phases
$\psi \to p\bar{p}$, \, $\psi \to \Lambda\bar{\Lambda}$, \, $\psi \to \Sigma^0 \bar{\Sigma}^0$, \,
$\psi \to \Sigma^+ \bar{\Sigma}^-$ and $\psi \to \Xi^0 \bar{\Xi}^0$ are large ($\phi_\psi \sim 200^\circ$)
and can be measured accurately in these channels.
It should be noted that the generation of a large phase for the production of a hyperon pair in the $e^-e^+$ annihilation
process was shown in the present work and in a series of other works \cite{Ahmadov1,Bystritskiy,BA,Ahmadov2,Ahmadov3,Kuraev}.
Finally, in conclusion, I report that in the future I plan also to consider other binary processes of final
state formation activated by charmonium annihilation.

\section{Acknowledgements}
I am grateful to Yu.M. Bystritskiy, M.A. Ivanov and V.I. Zakharov for useful discussions.

\section{Data Availability}
No data were created or analyzed in this study.

\end{document}